\documentstyle[12pt,aaspp4]{article}
\begin{document}
\def \m {$ M_\odot$~}
\def \l {$ L_\odot$}
\def \ro {g/cm$^{3}$}
\def \etal {\it et al.\rm}
\title{ The Sensitivity of Multidimensional Nova Calculations to the outer Boundary Condition }
\author{ S. Ami Glasner and Eli Livne}
\affil{Racah Institute of Physics, The Hebrew University, Jerusalem,  Israel}
\author{ James W. Truran}
\affil{Laboratory for Astrophysics and Space Research,
Enrico Fermi Institute, University of Chicago, Chicago, IL 60637, USA}
\begin{abstract}
 Multidimensional reactive flow models of accreted hydrogen rich envelopes
on top of degenerate cold white dwarfs are very effective tools for the study
of critical, non spherically symmetric, behaviors during the early stages of nova
outbursts. Such models can shed light both on the mechanism responsible for the
heavy element enrichment observed to characterize nova envelope matter and on the
role of perturbations during the early stages of ignition of the runaway.
The complexity of convective reactive flow in multi-dimensions makes the
computational model itself complex and sensitive to the details of the numerics.
In this study, we demonstrate that the imposed outer boundary condition can have
a dramatic effect on the solution. Several commonly used choices for the outer boundary
conditions are examined. It is shown that the solutions obtained from Lagrangian
simulations, where the envelope is allowed to expand and mass is being conserved,
are consistent with spherically symmetric solutions.
In Eulerian schemes which utilize an outer boundary condition of free outflow,
the outburst can be artificially quenched.
\end{abstract}
\section {Introduction}
The standard model for nova outbursts is a thermonuclear runaway in
an accreted hydrogen envelope on the surface of a white dwarf.
On time scales of days prior to the runaway, the flow in the partially degenerate
hydrogen rich envelope is convectively unstable.
The convective instability is driven by nuclear burning that  is highly sensitive
both to the temperature and to the chemical abundances. The details of the abundance mixing
within and at the edge of the convective zone and the fate of temperature fluctuations
are therefore crucial elements of the runaway mechanism. In order to capture these
effects, several multidimensional simulations of nova outbursts have been performed
(\cite{saf92}, \cite{sha94}, \cite{gl95}, \cite{glt97},
\cite{Ker2D}, \cite{Ker3D}, \cite{dursi02},\cite{alexakis04}).

In this study we attempt to validate multi dimensional schemes by discriminating
numerical effects from physical ones in the explosion phase.
One problem is to identify proper initial conditions for the simulations.
Since the pre-runaway quasi-hydrostatic accretion time is orders of magnitude
longer than the runaway time, all of the published multi dimensional nova
calculations start from a spherically symmetric 1D model, mapped to 2D (or 3D).
The computed measurable variables are sensitive to the 
adaptation of the 1D initial model to the multidimensional model, a problem that
was discussed by (\cite{Ker2D}).

Another problem is the implementation of a correct
outer boundary condition.
\cite{nariai80} discuss the sensitivity of 1D Lagrangian models to the outer boundary 
conditions during the expansion phase. They claim that fine zones are needed 
in order to resolve the details of the expansion of the nova atmosphere from white 
dwarf dimensions to red giant dimensions. 

Two independent studies - those of (\cite{gl95}, \cite{glt97}) and (\cite{Ker2D};
\cite{Ker3D}) - used the same 1D initial models
but were led to different conclusions about the strength of the runaway and its 
ability to reproduce a fast nova. In order to identify the reasons for these
differences, we carefully analyzed the sensitivities mentioned above.
 We argue here that the early stages of the flash phase, for which the evolution 
is still almost quasi-static, are highly sensitive to the outer boundary conditions. 
The sensitivity addressed in this study takes place during the  explosion (flash) phase  
and the main issue is the difference between multi-dimensional numerical schemes 
(Lagrangian vs. Eulerian), which has no 1D counterpart.
The sensitivity is certainly not a problem of resolution (see next section).
In the explosion (flash), phase the issue of resolution is important mainly in the
mixing layers, but not outside at the atmosphere.
 In this phase, as the flash develops,
the radial dimension of the envelope can increase by a factor of three (Fig.3),
affecting the pressure at the bottom of the accreted envelope where burning takes place.

The question of proper boundary conditions for multi dimensional simulations during this
phase is therefore, the main topic addressed in this work. 
We show that different, commonly used, mass outflow conditions at the outer
boundary can make the difference between
a successful runaway model and a model for which the runaway is artificially
quenched. We analyze the cause for this failure and conclude with a recommendation
for boundary conditions that are consistent with the physics of the problem.

\section {Problematic outer boundary conditions - which and why}

Our initial 1D model is a 1.14 \m CO WD that has accreted a $ 2.6 \times 10^ {-5} $ \m 
envelope (solar abundance) up to the stage at which a TNR develops. 
The model is composed of 130 CO core zones and 50 accreted hydrogen zones.
In 1D, the main features of the runaway, such as the accreted mass up to the runaway,
the peak temperature and the peak energy generation rate, are converged at that
spatial resolution. Increasing the number of zones in the accreted hydrogen
layer to 250 change those values by a fraction of a percent.

When the temperature at the base of the envelope has reached
$1 \times 10^{8} $ K (just before the 1D runaway starts),   the outer 
$ 6 \times 10^ {-5} $ \m  of the star is mapped, without any change in the zoning, 
as a multidimensional model (2D or 3D) onto the multidimensional
hydro solver. The evolution in 2D is followed without any initial perturbation.
The 1D code and the 2D code use the same difference scheme for the momentum equation,
the same radial zoning and the same EOS. 
Therefore,  after the initial mapping the envelope is almost everywhere in full hydrostatic 
equilibrium. Deviations from hydrostatic equilibrium at local points are less
then one part in ten thousand. For models at earlier stages in the runaway where the
envelope is still stable against convection, the 2D mapped model stays in perfect
hydrostatic equilibrium for a hundred seconds (the typical time step, dictated
by the sound crossing time is  about $  10^{-3} $ seconds). In this way we avoid the
problems of mapping the 1D models to multi dimensional models, a problem
that is thoroughly discussed by (\cite{zingale02}).

The numerical scheme of the 1D model is a Lagrangian scheme for which mass at the 
outer boundary is conserved and there is no overshoot mixing between the core matter
and the accreted envelope. The multidimensional models try to analyze closely the
details of the convective flow and are therefore basically Eulerian.
Because of the extreme effect that overshoot mixing at the inner interface has on the
nuclear burning in 2D it is impossible to judge the accuracy of the 2D model by the 1D results.

 The models presented here are all 2D models.
 The 2D hydro VULCAN solver (\cite{liv93} consists of two computational stages within one time step. 
 As a first stage a purely
 Lagrangian hydro time step is performed (explicit in our case). In the second stage , a high-order
 explicit remapping is performed. In this remapping stage one can remap the new flow field
 onto any desired grid; the original Eulerian grid, or any other adaptive grid. The scheme as a whole
 is therefore called hereafter Arbitrary Lagrangian Eulerian (ALE).   
 The lateral boundary conditions at the sides of 
 the multidimensional slice can be either fixed or periodic with only a slight 
 effect on the solution. 

 The computational slice includes a significant part of the static, inert, 
 white dwarf core. The deep cold part of the core justifies a fixed inner 
 boundary condition. We assumed initially 
 that, since the computational slice covers a few pressure scale heights, there will
 not be any sensitivity to the exact outer boundary condition. 
 Subsequently, we determined that this assumption
 is quite unjustified. It is well known that the runaway is a threshold phenomena,
 the pressure at the base of the envelope must be in excess of a certain value
 of a few time  $ 10^{19}  erg/cm^{3} $ (\cite{Fujimoto82}, \cite{macdonald83}).
 Spurious mass loss 
 from the computational grid at the outer boundary slightly decreases
 the pressure at the bottom of the burning zone. In the worst case, this mass loss 
 totally quenches the runaway while in other cases it leads to a very fast 
 non-physical turnoff of the runaway.
 Since the physical mechanism for the turnoff is  the decrease in pressure
 due to expansion, it is clear that a correct boundary condition is essential.
 In order to demonstrate the problem and its solution, we present here the results
 we have obtained for four
 different schemes for the outer boundary condition that were tested:

\placefigure{Fig1}

\begin {itemize}

\item  Eulerian scheme with free outflow at the outer boundary 
(Eulerian - open in Fig. 1):
  For this scheme, matter with positive radial velocity in the outer zone
  is lost from the computational grid at each step.

\item  Eulerian scheme with  modified mass flux outer boundary condition 
(Eulerian inflow+outflow in Fig. 1):
This scheme derives from the observation that during the development of 
the runaway the outer boundary
is convective. It follows that at any time some of the 
outer grid points have velocities directed outward and some have incoming
velocities. Therefore, as long as the evolution is quasi-static,
the amount of mass 
that leaves the last original zone
should flow back onto the grid once the convective turnover of this convection 
cell is 
completed. Based on this assumption, we added a modified mass flux outer boundary
condition. For this scheme, whenever the velocities in the outer zones point
outward, the mass flux is calculated and mass is lost. When the velocities 
point inward, however, we impose an incoming mass flux with the assumption that
the outer zone is also the donor. This method is in a way equivalent to models
working with ``ghost cells'' for extrapolation of the mass fluxes in the case of 
``open boundaries''.

\item   Eulerian scheme with no mass flux (fixed) outer boundary condition 
(Eulerian fix - closed in Fig. 1):
For this option, matter is forced to stay within the grid (no mass flux is allowed to 
pass into or out of the outer edge of the grid). 
Mass that is artificially accumulated near the outer boundary produces extra outer 
pressure. This extra pressure is only an artifact of the numerical model.

\item  An Arbitrary Lagrangian Eulerian (ALE) scheme for which the 
radius of the outer zone is defined in a way that conserves mass 
(Lagrangian in Fig. 1). This scheme enables us to overcome three major obstacles
that the nova models face:
 a) The Lagrangian hydro step takes care of the dimensional expansion of the computed
 regime.
 b) The remap stage enables us to keep a radial grid  all the time.
    The radial grid has $K\_max$ zones. The number of radial zones is fixed and they
    are numbered by the index $k$, $k=1,K\_max$ . The transversal zones are numbered by
    the index j, $j=1,J\_max$. All the transversal zones ($j=1,J\_max$) for each k
    are defined to be the k-th shell.
 c) The arbitrariness of the Eulerian grid enables us to model the burning zones
    at the bottom of the hydrogen rich envelope with very delicate zones in spite of the
    dimensional expansion due to the runaway.

  In order to realize these goals 
 at each time step, a ``radial''
 mass conserving grid is defined ( after the Lagrangian part of the step) in the following way. 
 For each k shell of grid cells 
 that started the step with the same radius a new radius 
 is defined so that the mass of that shell is conserved. In the new grid, defined 
 for the next step, the lateral
 boundaries of each grid cell are kept fixed ($J\_max$ lateral equal zones).
 The radial coordinate of the whole k-th shell of grid cells is defined by interpolation 
 between the initial Eulerian radial grid and the new mass conserving grid. 
 The only restriction is that, at the outer parts
 of the grid, the $K\_max$ shell will follow the
 new mass conserving grid so that the total mass is conserved.
 Since we usually want to keep the fine zoning in the inner parts of the
 grid where intense burning and mixing takes place ( item (c) above), the new
 grid
 smoothly interpolates between the initial Eulerian grid in the inner zones and the
 ``Lagrangian'' grid at the outer zones. In this way the outer boundary is
 expanding with time. 
 The interpolation we use is of the following form:

\begin{equation}
 R_{new}(k)=f\times R_{Lag}(k)+(1-f)\times R_0(k);   \ \ \ \ \ \ \   f=(k/K_max)^n
\end{equation}

     where: 
\begin {itemize}
            \item
            $k$ is the index of the radial k-th shell.
            \item 
            $R_{Lag}(k)$ is the radius of the k-th shell in the mass conserving Lagrangian grid.
            \item 
            $R_0(k)$ is the original radius of the k-th shell in the initial Eulerian grid
            \item 
            $n$ is a power low. The default value we use is $n=3$.
\end {itemize}
\end {itemize}

In order to avoid small differences that might appear during the initial stages, 
(just after the mapping of the 1D model into 2D), 
all four models started from a single 2D profile of the ALE scheme 20 seconds 
after it was mapped into 2D.
The convective cells are already fully developed at this stage. 

 With the present 
resolution, it is difficult to obtain any significant information on the light curve. 
The results concerning the strength of the outburst are therefore judged only by 
comparison with 1D spherically symmetric models
(\cite{nariai80} ; \cite{pk84} ; \cite{sst85} ; \cite{sst86} ; \cite{pri86}). 
In this respect, we define a successful nova to be a TNR for 
which the amount of thermonuclear energy pumped into the envelope 
on a dynamic time scale is comparable to the binding energy of the envelope. 
For the case we have considered of a 1.14 \m  WD with an accreted envelope mass of 
$ 2.6 \times 10^{-5} $ \m, the binding energy of the envelope 
is approximately $  10^{46} $ erg.
The criterion we therefore adopt for a successful nova is an overall
reaction rate greater 
then $ 1.0 \times 10^{44} $ erg/sec for more then 50 seconds (typical
 dynamical time of the accreted envelope).
We examine the sensitivity of the 2D evolution of the models 
by comparing the following measurable quantities;

\begin {itemize}

\item
 The time history of the overall energy production rate (Fig. 1).
\item
 The time history of the pressure at the base of the computational slice (Fig. 2).
  
\end {itemize}

\placefigure{Fig2}

The ALE scheme is expected to be the most accurate. Indeed, for this scheme 
the energy production rate increases continuously
and a runaway is taking place. The characteristics of this model throughout its 
evolution is compatible with the 1D evolution of CO enriched models.
We have therefore chosen to compare all other models to this model.

The free Eulerian-open  scheme gives a significantly different result: 
essentially no runaway. Although the initial pressure at the  
outer zone is two orders of magnitude less than the pressure at the bottom zone, 
the small decrease in pressure on a few dynamic times 
(the flow is  extremely sub-sonic) quenches the runaway.
The differences in the pressure profiles are evident in Fig.2. 
In Fig.3 we demonstrate the major difference between the two schemes. The figure
shows a color map of the logarithm of the energy production rate within the envelope
close to the maximum of energy production (time=100 sec, see Fig. 1) in both models.
For the ALE Lagrangian mass conserving scheme we see the high energy production
rates at the bottom, reaching values of ${1-2}\times 10^{17}$ erg/gr/sec. For the pure Eulerian 
scheme the rates are below $1 \times 10^{15}$ erg/gr/sec. The differences in the geometric
dimensions of the envelope in both models is clear, once observed it confirms the necessity
of the Lagrangian approach. 

The modified Eulerian scheme conserves mass for a while, 
as long as the matter is convecting in the absence of overall expansion.
Once the runaway starts to take place (at about $t=80$ seconds),  however, 
the net expansion velocities in the outer regions are higher than
any component of the convective flow at this zone. 
Subsequently, mass loss quenches the runaway (Fig. 1).

The Eulerian-closed scheme agrees quite well with the ALE scheme until 
the point at which  pressure starts to drop
in the Lagrangian scheme ($t=180$ seconds in Fig. 2). The agreement is both from the 
point of view of energy production rates (Fig. 1) and of the topology of the 
convective cells. Once we prevent the expansion of
matter at the outer boundary, however, it flows back and artificially enhances the
convective velocities and the mixing (Fig. 2).
This scheme thus extends the runaway artificially so that
in the absence of expansion there will be no turnoff of the TNR.

\section {Discussion and Conclusion}

The onset of the thermonuclear runaway in nova outbursts occurs 
once the accreted envelope reaches a critical pressure. 
A detailed knowledge of the evolution of the pressure 
at the base of the envelope is therefore essential to our understanding 
of the development of the runaway. The numerical simulations presented in this
paper show the major role which the outer boundary conditions play in
the evolution of the outburst.
Of the four schemes we examine in this paper, only the ALE scheme 
provides a correct physical solution to the problem.
The free outflow (open) boundary condition used in Eulerian schemes 
artificially quenches the runaway.
Use of such a boundary condition in Eulerian simulations 
can explain the main differences between the models presented by
\cite{glt97} and those of \cite{Ker2D}.
The fact that in the latter simulation virtually 
all of the hydrogen disappears from the grid at late times 
(Fig.8 in \cite{Ker2D}) supports this assumption (Fig.3).

For the same reason, the modified boundary condition we have described fails to
reproduce the physical features of nova outbursts during the expansion phase.
It can serve properly to represent the boundary conditions only in the case of
a quasi static convective envelope. 

A possible remedy to this problem could be an  addition of empty 
grid cells at the outer part of the computed region (``ghost cells'').
The amount of expansion demonstrated in Fig.3 shows that this solution
is not appropriate for the problem presented here.

To conclude, we state that proper multi dimensional simulations of Novae-like
explosions should maintain the Lagrangian nature of the expanding envelope.
 The sensitivity we encountered is expected whenever the onset of a runaway
is critically dependent upon the evolution of the pressure in a subsonic regime.

\section*{Acknowledgments}
This work is supported in part at the University of Chicago (JWT) by the
U.S. Department of Energy, under Grant B523820 to the ASCI Alliances 
Center for Astrophysical Flashes and Grant DE-FG02-91ER40606 in 
Nuclear Physics and Astrophysics and by the NSF under grant PHY 02-16783
for the Physics Frontier Center ``Joint Institute for Nuclear Astrophysics.''
Ami Glasner, wants to thank the Department of Astronomy and Astrophysics, 
University of Chicago and the members of the ASCI Center for Astrophysical
Thermonuclear Flashes for their hospitality during his visits to Chicago, 
where part of this work was done.

\clearpage

\figcaption[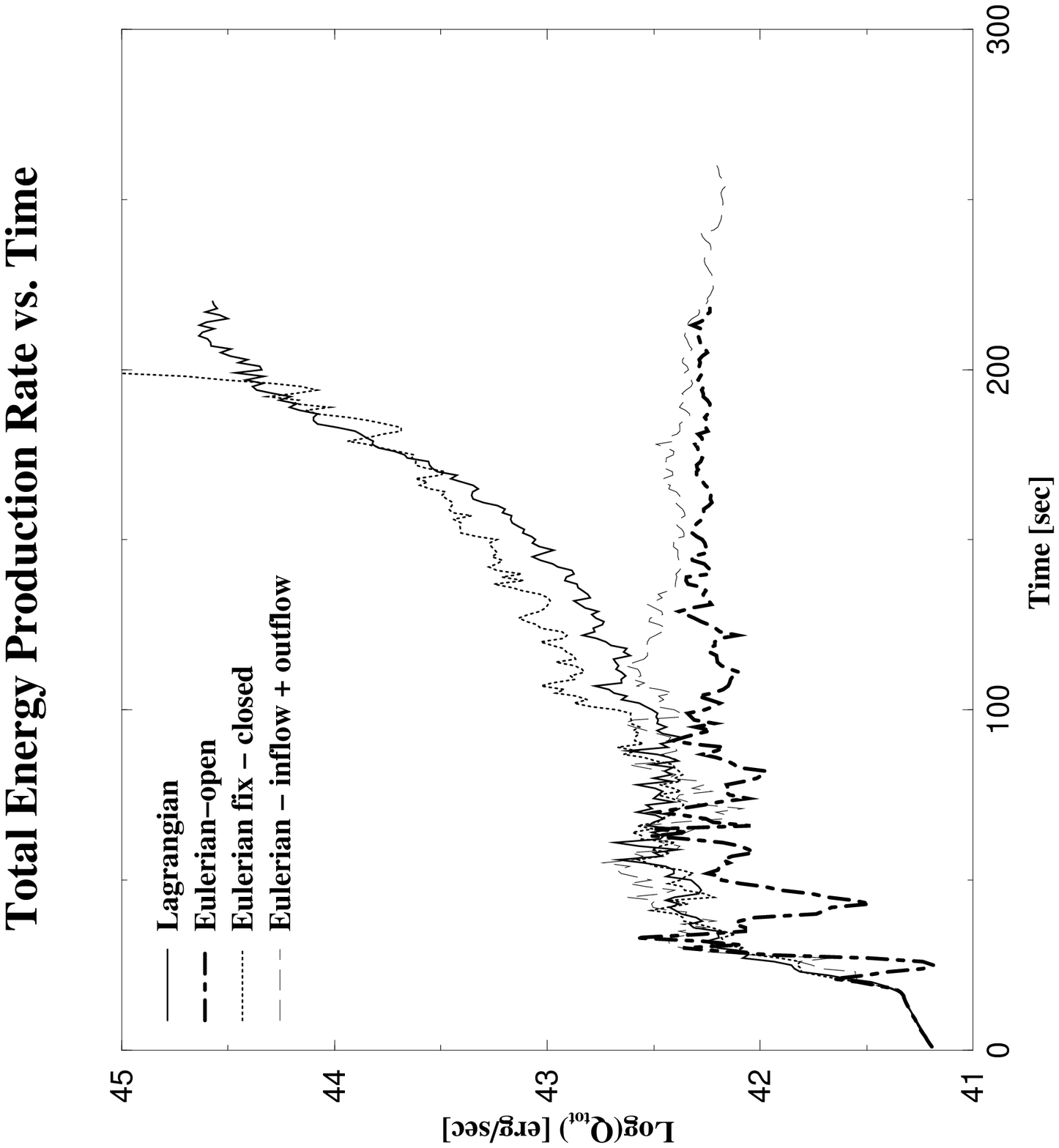]{Total burning rate as a function of time for all models (see text)\label{Fig1}}

\figcaption[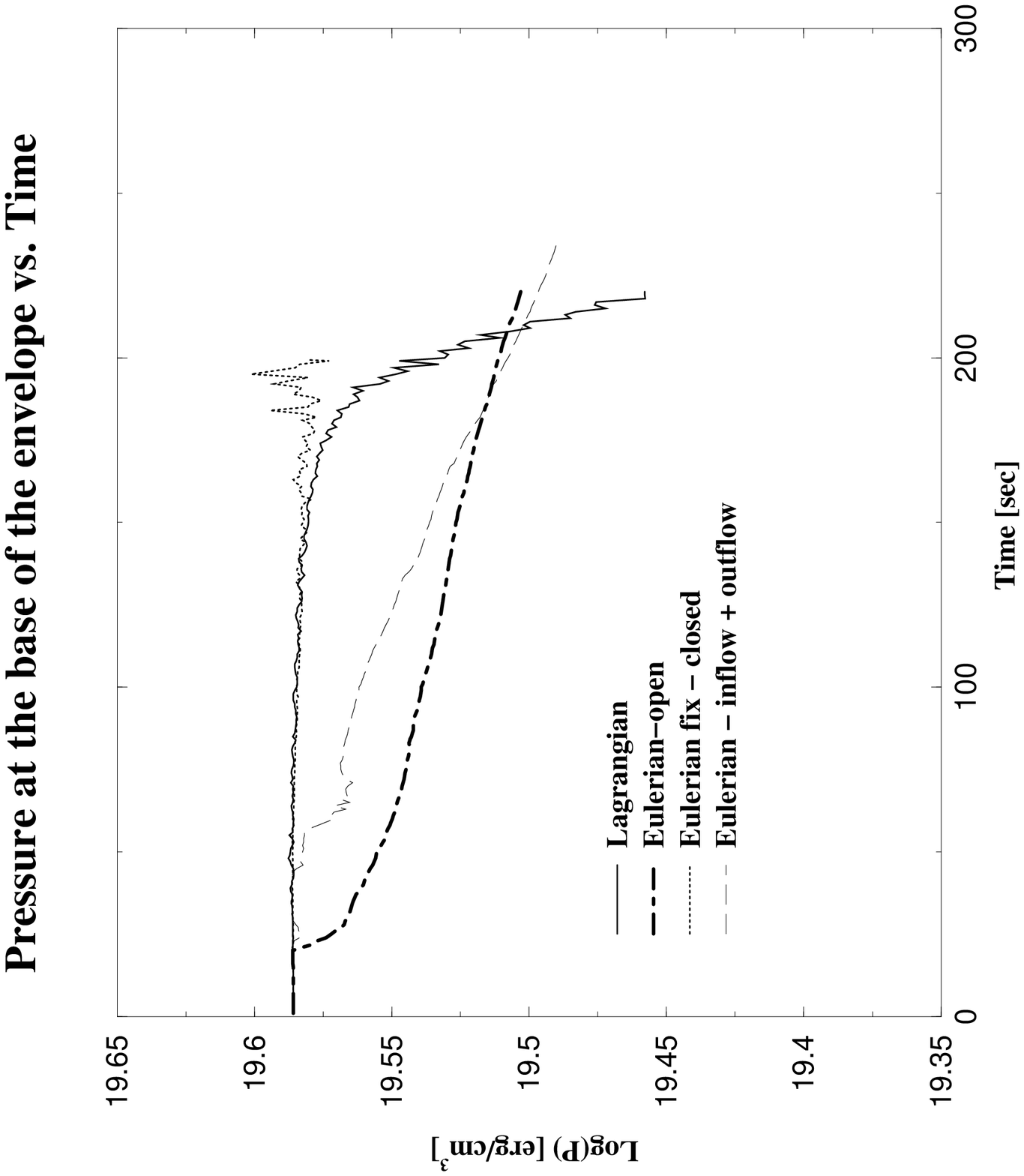]{Pressure at the base of the envelope as a function of time for 
           all models (see text) \label{Fig2}}

\figcaption[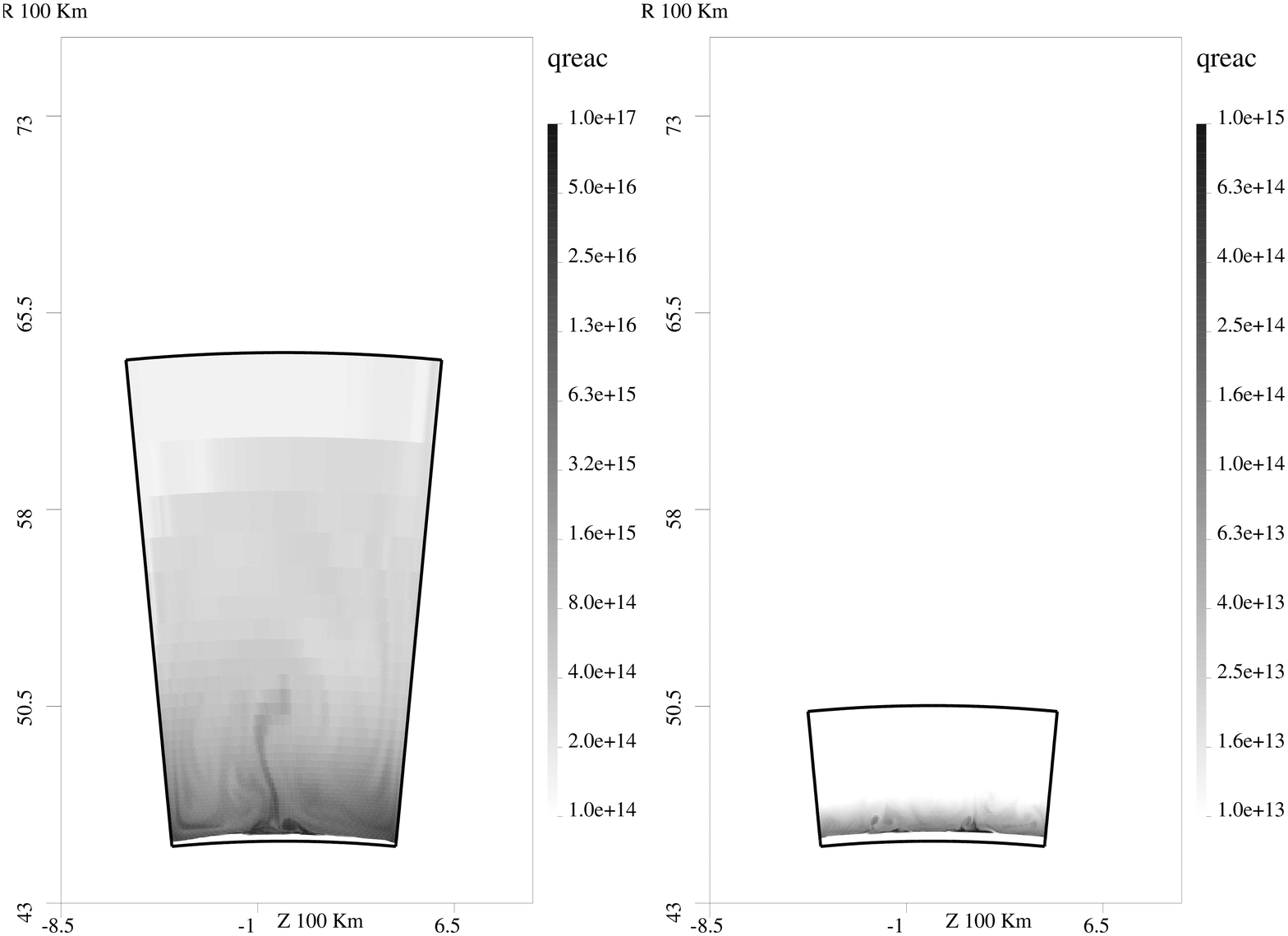]{Color map of the thermonuclear energy production rate at time=100 sec for the 
           pure Eulerian 
           case-right and the ALE Lagrangian scheme-left. The spatial coordinate is in units of 100 Km. 
           The energy production rate is in erg/gr/sec. The rate scale is different in the two
           cases (see scale to the right of each model) (see text) \label{Fig3}}

\newpage
\plotone{f1.eps}
\newpage
\plotone{f2.eps}
\newpage
\plotone{f3.eps}

\end{document}